\documentclass[pra,aps,twocolumn,superscriptaddress,notitlepage]{revtex4-2}

\usepackage{amssymb,amsfonts,amsmath,amstext,graphicx,hyperref,lipsum,empheq}
\usepackage{float}
\usepackage[normalem]{ulem}
\usepackage{hyperref}
\hypersetup{
    colorlinks=true,
    linkcolor=blue,
    citecolor=blue,  
    filecolor=magenta,
    urlcolor=cyan,
}
\usepackage{outlines}
\usepackage[dvipsnames]{xcolor}
\usepackage{hyperref}

\usepackage{subcaption}
\usepackage{ragged2e}
\usepackage{bm}

\newcommand{\eps}[0]{\epsilon}
\newcommand{\beq}[0]{\begin{equation}}
\newcommand{\eeq}[0]{\end{equation}}
\newcommand{\non}{\nonumber}
\def\be{\begin{equation}}
\def\ee{\end{equation}}
\def\bea{\begin{eqnarray}}
\def\eea{\end{eqnarray}}
\newcommand{\ba}{\begin{eqnarray}}
\newcommand{\ea}{\end{eqnarray}}

\def\BraVert{\egroup\,\mid\,\bgroup}

\definecolor{myblue}{rgb}{.8, .8, 1}

\makeatletter
\newenvironment{panelTR}[2]{%
  \begin{minipage}[t]{#1}%
  \raggedright 
  \def\@tempa{#2}%
  \ifx\@tempa\@empty
  \else
    \makebox[0pt][l]{\raisebox{-10.5ex}{\hspace{10ex}\textbf{(#2)}}}%
  \fi
}{%
  \end{minipage}
}
\makeatother

\begin{document}
\nocite{apsrev42Control}

\title{Quantum sensing with discrete time crystals in the Lipkin-Meshkov-Glick Model}

\date{\today}
\author{Rahul Ghosh}
\affiliation{Department of Physical Sciences, Indian Institute of Science Education and Research Berhampur, Berhampur 760010, India}
\author{Bandita Das}
\affiliation{Department of Physical Sciences, Indian Institute of Science Education and Research Berhampur, Berhampur 760010, India}
\author{Victor Mukherjee}%
\affiliation{Department of Physical Sciences, Indian Institute of Science Education and Research Berhampur, Berhampur 760010, India}

\begin{abstract}
Quantum phase transitions have been shown to be highly beneficial for quantum sensing, owing to diverging quantum Fisher information close to criticality. In this work we consider a periodically modulated Lipkin-Meshkov-Glick  model to show that discrete time crystal (DTC) phase transition in this setup can enable us to achieve quantum-enhanced high-precision  sensing of field strength. We employ a detailed finite-size scaling analysis, a time-averaged Inverse Participation Ratio analysis, and mean-field analysis in the thermodynamic limit, to determine the critical properties of this second-order phase transition. Our studies provide a comprehensive understanding of how quantum criticality in DTCs involving long-range interactions can be harnessed for advanced quantum sensing applications. 
\end{abstract}

\maketitle

\section{Introduction}
\label{sec I intro}
In the recent years there has been extensive research on harnessing quantum systems for devising high-performing quantum technologies, including quantum heat engines and refrigerators \cite{bhattacharjee20quantumreview, campbell25roadmap}, quantum sensors \cite{correa15individual, ilias2022_critical, sarkar2025_exponentially, agarwal2025_quantum}, and quantum thermal transistors \cite{joulain16quantum, gupt22floquet}. The possibility of realizing these technologies in currently existing platforms has significantly enhanced the importance of theoretical studies of these technologies. Among these, quantum sensors, such as quantum thermometers \cite{hofer17quantum, mehboudi19thermometry} and quantum magnetometers \cite{brask15improved, bhattacharjee20quantum} have been shown to be significantly beneficial for high-precision estimation of different parameters.  The precision of a quantum sensor can be quantified by the quantum Fisher information (QFI), which signifies the rate of change of the system state for a small change in the parameter to be estimated \cite{paris09quantum}. Consequently, the importance of  studies of the behavior of QFI in quantum systems, in particular its scaling with system size $N$, cannot be overstated. In this respect, the Standard Quantum Limit (SQL) allows QFI to  scale at most linearly with system size \cite{giovannetti04quantum}. Therefore harnessing quantum properties of materials to model quantum sensors which can overcome the SQL, thereby exhibiting superlinear scaling of QFI, is one of the major aims in the field of quantum sensors. In this respect, recent works have highlighted the important role played by quantum phase transitions in engineering high-precision quantum sensors. Non-trivial changes in the qualitative nature of the ground state close to criticality leads to divergence in various parameters, including QFI. Researchers have harnessed this unique feature of quantum critical systems to design  quantum sensors exhibiting super-linear scaling of QFI close to criticality \cite{garbe22critical, ilias2022_critical, sarkar2025_exponentially, beaulieu25criticality}. More recently, a separate class of phase transitions in quantum systems driven out of equilibrium, viz. discrete and boundary time crystals, which are associated with time-translational symmetry breaking in many-body systems, have been shown to be immensely beneficial for  engineering high-performing quantum sensors \cite{montenegro23quantum,  iemini24floquet, cabot24continuous, yousefjani25discrete}. 

Time-crystals are a relatively recently explored non-equilibrium phase of matter, and are associated with long-range order in time and space \cite{zaletel23colloquium, else20discrete}. Boundary or continuous time crystals have been shown to emerge in the presence of dissipation \cite{iemini18boundary}, while discrete time crystals are generated through periodic modulation in the system Hamiltonian, and have been realized in disordered systems \cite{yao17discrete},  systems with stark localization \cite{liu23discrete}, in the presence of non-Hermitian Hamiltonians \cite{yousefjani25non}, in the presence of unbounded potentials \cite{barlev24discrete}, in central-spin models \cite{biswas25discrete} and in systems with long-range interactions \cite{russomanno17floquet, mishra2025_coexistence}. Relatedly, long-range interactions in the form of spin string operators have also been shown to support spontaneous breaking of continuous time translation symmetry  in the presence of unitary dynamics~\cite{kozin2019_quantum}. Beyond its fundamental importance for understanding the physics and thermodynamics of quantum systems driven out of equilibrium \cite{paulino23nonequilibrium}, time crystals have also been shown to be helpful for designing various quantum technologies, including quantum engines \cite{carollo20nonequilibrium}, quantum sensors \cite{iemini24floquet, yousefjani25discrete, sahoo2025_power, shukla2025_prethermal} and quantum clocks \cite{viotti25quantum}, and have been realized experimentally using optical cavity setups \cite{kongkhambut22observation, taheri22all} quantum processors \cite{mi22time}, ion traps \cite{zhang17a}, in various mesoscopic setup \cite{autti2022_nonlinear, autti2021_ac}  and Rydberg gases \cite{wu24the}.  Importantly, time crystals are robust phases of matter, which do not need fine tuning of the system parameters for its existence. This particular property of time crystals can be immensely advantageous for designing quantum sensors, which under practical scenarios, may need to work in the presence of different impurities.  
In this work  we focus on quantum sensors based on the Lipkin-Meshkov-Glick (LMG) model \cite{ribeiro2007_thermodynamical, ribeiro2008_exact, dusuel2004_finite} driven out of equilibrium. As shown in Ref. \cite{russomanno17floquet}, this model exhibits a discrete time crystal (DTC) phase for periodic spin flip modulations.  We use an order parameter, susceptibility, QFI and inverse participation ratio to study the critical properties associated with the DTC to non-DTC phase transitions, and show that this phase transition is associated with a quantum-enhanced superlinear scaling of QFI with $N$. We apply finite-size scaling analysis to investigate the universal scaling forms associated with an order parameter and QFI, present critical exponents associated with this second order phase transition,  and compare our results with those obtained through mean-field analysis in the thermodynamic limit. Finally, we also present dynamical phase diagrams w.r.t. an impurity parameter $\epsilon$, transverse field $h$ and the time period of modulation $\tau$.

The paper is organized as follows: in Sec. \ref{secI} we briefly introduce the concept of QFI, and it's role in quantum sensing; in Sec. \ref{secIII} we discuss the model and dynamics; we analyze the behavior of the order parameter and the associated susceptibility in Sec. \ref{secIV}; study QFI and quantum sensing in Sec. \ref{secV}; study the DTC to non-DTC phase transition using the time-averaged inverse participation ratio in Sec. \ref{secVI}; show the phase diagram of the model with respect to transverse magnetic field, time period of modulation and an error parameter $\epsilon$ in Sec. \ref{secVII}. We finally conclude in Sec. \ref{secVIII}.

\section{Quantum parameter estimation}
\label{secI}
In this study, we address the estimation of a single parameter, $\epsilon$, through measurements on a quantum state $\rho(\epsilon)$, which encodes the parameter. The quantum state, called the probe state, is utilized to perform measurements, with the resulting outcomes processed by an estimator function to deduce the value of the parameter accurately. Measurements are typically represented by a set of Positive Operator Valued Measurements (POVM) $\{\Pi_n\}$, where the probability of the $n$-th outcome is given by $p_n(\epsilon) = \text{Tr} \left( \rho(\epsilon) \Pi_n \right)$. The uncertainty in estimating $\epsilon$, quantified as the standard deviation $\delta\epsilon$, is governed by the Cramér-Rao inequality $\delta\epsilon \geq \frac{1}{\sqrt{\mathcal{M} \mathcal{F}_C}}$. Here, $\mathcal{M}$ signifies the total number of measurements, and the classical Fisher information (CFI) is described by $\mathcal{F}_C = \sum_n p_n \left( \partial_\epsilon \log p_n \right)^2$ \cite{Qvarfort2018_gravimetry,yousefjani23floquet, montenegro24quantum, sarkar2025_exponentially}. By maximizing the CFI over all potential measurements, we derive the QFI $\mathcal{F}_Q = \max_{\{\Pi_n\}} \mathcal{F}_C$, a measurement-independent metric. Consequently, the Cramér-Rao inequality is refined to
\begin{equation}
\delta\epsilon \geq \frac{1}{\sqrt{\mathcal{M}\mathcal{F}_C}} \geq \frac{1}{\sqrt{\mathcal{M}\mathcal{F}_Q}},
\end{equation}
where the QFI sets the ultimate precision benchmark for the estimation. Intriguingly, the evaluation of the QFI circumvents the cumbersome optimization over all measurement bases by employing the symmetric logarithmic derivative (SLD) operator $\mathcal{L}$, defined as:
\begin{equation}
\frac{\partial \rho(\epsilon)}{\partial \epsilon} = \frac{\rho(\epsilon)\mathcal{L}_\epsilon + \mathcal{L}_\epsilon\rho(\epsilon)}{2}.
\end{equation}

The QFI can thus be expressed in terms of $\mathcal{L}$ as $\mathcal{F}_Q = \text{Tr} \left(\rho(\epsilon)\mathcal{L}_\epsilon^2\right)$. For pure states, where $\rho(\epsilon) = |\psi(\epsilon) \rangle \langle \psi(\epsilon)|$, the expression simplifies to $\mathcal{L}_\epsilon = 2\partial_\epsilon \rho(\epsilon)$, and consequently:
\begin{equation}
\mathcal{F}_Q(\epsilon) = 4 \left( \langle \partial_\epsilon \psi(\epsilon) | \partial_\epsilon \psi(\epsilon) \rangle - \left| \langle \partial_\epsilon \psi(\epsilon) | \psi(\epsilon) \rangle \right|^2 \right),
\label{eqQFI}
\end{equation}
where $|\partial_\epsilon \psi(\epsilon) \rangle$  denotes the derivative of the state  $|\psi(\epsilon) \rangle$  with respect to the parameter $\epsilon$ (see appendix \ref{sec Appendix A}). Since the QFI quantifies the rate of change of the probe state, it equates to the fidelity susceptibility \cite{garbe2022_critical}. To meet the ultimate precision limit specified by the QFI, measurements must occur in the optimal basis, determined by the projectors derived from the eigenvectors of the SLD operator $\mathcal{L}_\epsilon$, although this optimal basis is not unique.

\section{LMG Hamiltonian and the Floquet protocol}
\label{secIII}

We consider the LMG model describing $N$-qubits in the presence of all-to-all uniform interactions, and subjected to a transverse field, described by the Hamiltonian 
\begin{align} \label{eq:LMG}
H_{LMG} = -2 \frac{J}{N} S_{z}^{2} - 2 h S_{x}, ~~~~ 0 < t < \tau,
\end{align}
where the collective spin operators are $S_\alpha = \sum_i \sigma^\alpha_i /2$ with $\alpha = x,y,z$ and $\sigma^\alpha_i$ are the Pauli matrices. Here, $J$ denotes the overall (infinite-range) coupling strength, and $h$ denotes the strength of on-site potentials. Here we restrict ourselves to the spin sector with $S=N/2$.  The above model shows a $\mathbb{Z}_2$ symmetry breaking for  $J>h$, and  supports a DTC phase in the presence of a periodic modulation given by \cite{russomanno17floquet}
\begin{align} \label{eq: floquet hamiltonian}
H(t)= H_1 + \sum_n \delta(t-n\tau) H_2 \non
\\
H_{1} = H_{LMG} + \delta S_{z}, \quad  
H_{2} = \phi \, S_{x}, 
\end{align}
where $\delta \ll 1$ denotes the strength of a weak field of degeneracy lifting, which ensures proper initialisation in a localized ground state with negative values of $S_z$, and $H_{2}$ implements an instantaneous global rotation (kick) by an angle $\phi$ about the $x$-axis. For simplicity, we set $J=1$ and $\delta = 10^{-5}$. The choice $\phi = \pi$ generates persistent period-doubling oscillations - a hallmark of DTCs - when the system is prepared in a symmetry-broken ground state. The perfect $\pi$-kick swaps the two degenerate symmetry-broken ground states. However to study the stability of period doubling, we consider imperfect spin flips by choosing
\begin{align}
\phi=(1-\epsilon)\pi,
\end{align}
where $\epsilon$ quantifies the deviation from a perfect $\pi$-kick. Finally, we can write the one-period Floquet operator as
\begin{align}
U_{F}=e^{-iH_{2}}\,e^{-iH_{1}\tau},
\label{eqfloquet}
\end{align}
such that the state after $n$ periods is
\begin{align}
\lvert\psi_{n}\rangle = U_{F}^{\,n}\lvert\psi_{0}\rangle,
\label{eqpsin}
\end{align}
where $\lvert\psi_{0}\rangle$ is the initial state.
 We choose the initial state as 
the symmetry-broken ground state of the time-independent
LMG Hamiltonian.


 As discussed in \cite{russomanno17floquet}, the DTC phase persists  in the limit of $N \to \infty$, and is characterized by a single peak in the Fourier spectrum at $\omega_{\rm DTC} = \pi/T$, with the height of the Fourier peak converging to a constant for large $N$. On the other hand, finite size systems are associated with two Fourier peaks at frequencies $\omega_{\pm} = \pi/\tau \pm \theta(N)$, with $\theta(N)$ vanishing exponentially with system size.
Furthermore, this phase is robust to small imperfections in the Hamiltonian, quantified here by the parameter $\epsilon$. Notably, as we show below, the regime of $\epsilon$ supporting a DTC phase increases with decreasing $h/J$, i.e., with increasing inter-spin coupling strength $J$, thus emphasizing the non-trivial many-body nature of this phase \cite{pizzi2021_higher}.

\begin{figure}[t]
\centering
\includegraphics[width=\linewidth]{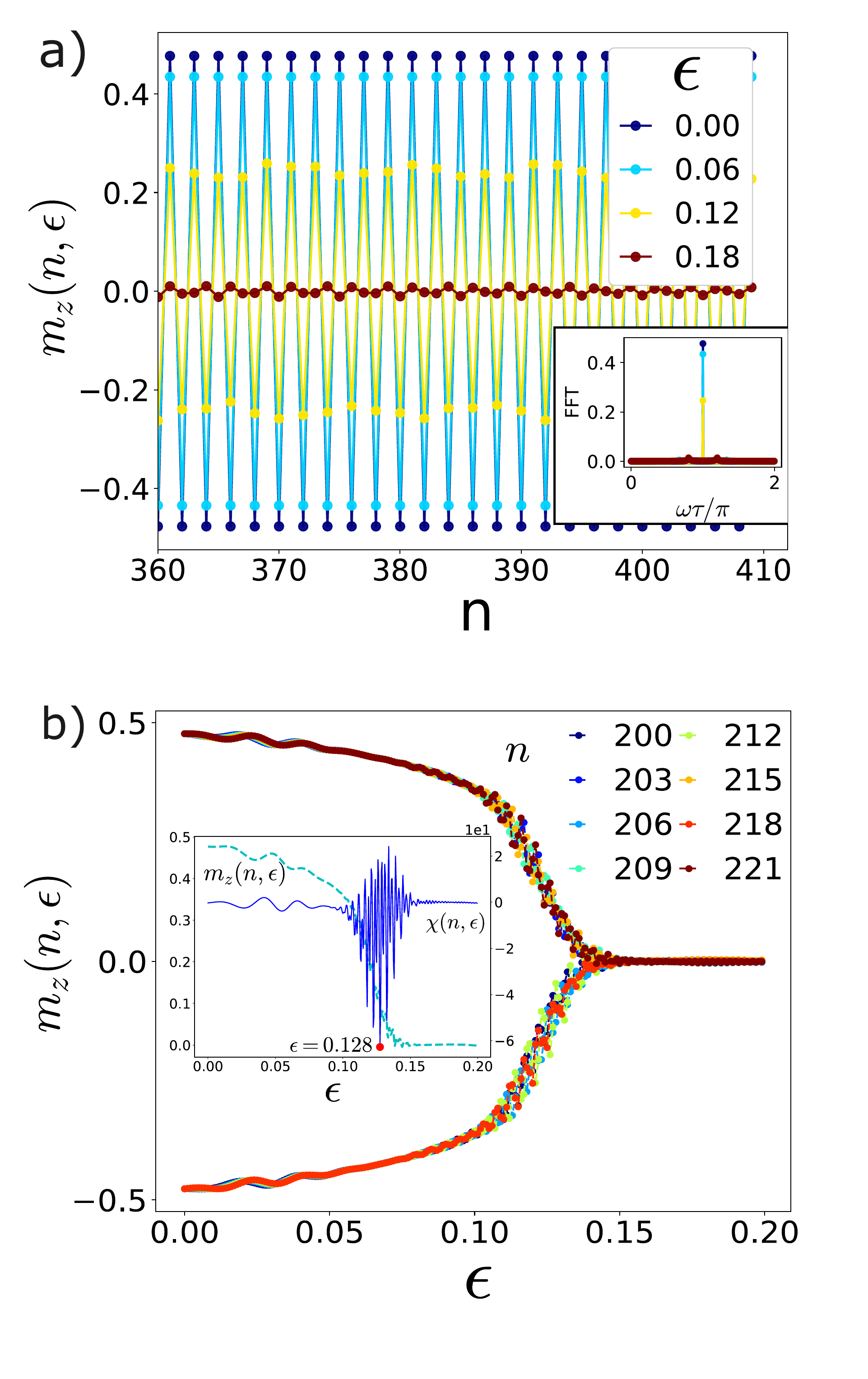}  
\caption{\justifying{
{\bf Magnetization and susceptibility:} 
(a) Stroboscopic magnetization $m_z(n, \epsilon)$ as a function of $\epsilon$ for $N = 1000$ over 400 periods. Inset: FFT of the magnetization signal. 
(b) Magnetization $m_z(n, \epsilon)$ shown for varying $n \in \{200, 203, \ldots, 221\}$ as a function of $\epsilon$.
Inset: plot of $m_z(n, \epsilon)$ and the susceptibility $\chi(n, \epsilon)$ versus $\epsilon$ for $n = 101$; the dip in susceptibility at $\epsilon = 0.128$ sets the critical imperfection threshold.
We have fixed $h = 0.3$ and $\tau = 0.6$. 
}}
\label{fig mag}
\end{figure} 

In the following section, we analyze the resulting discrete time-crystalline behavior by examining the magnetization dynamics along the $z$-direction and define a suitable order parameter.

\begin{figure}
\centering
\includegraphics[width=\linewidth]{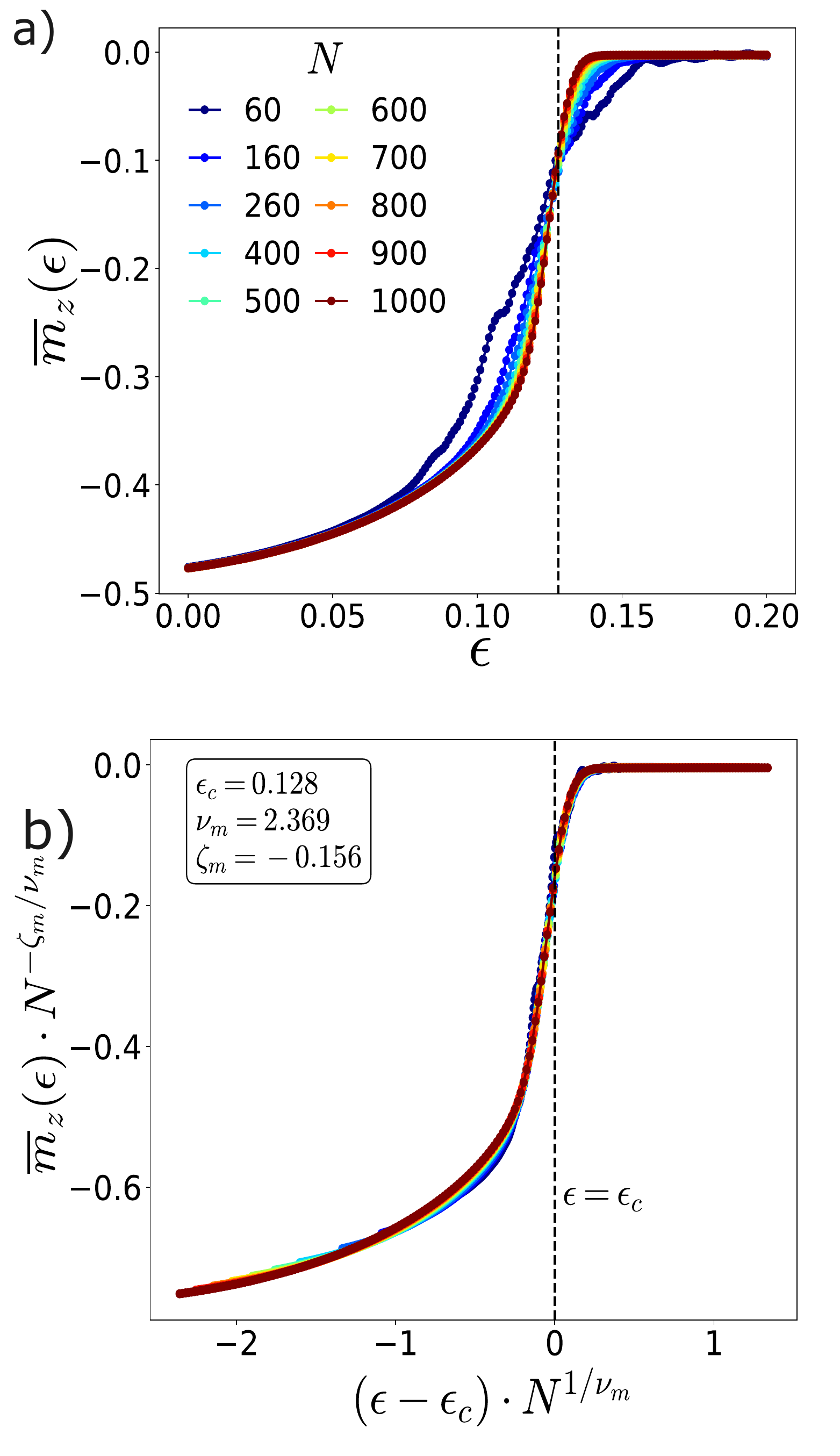}
\caption{\justifying{{\bf Order parameter:}
(a) Order parameter $\overline{m}_z(\epsilon)$ [cf. Eq.~\eqref{eqn order param}] versus $\epsilon$ for $N = 60$ to $N = 1000$.
(b) Finite-size scaling collapse: $\overline{m}_z(\epsilon) \cdot N^{-\zeta_m/\nu_m}$ versus $(\epsilon - \epsilon_c) \cdot N^{1/\nu_m}$. The value of the constants are $\nu_m = 2.369$, $\zeta_m = -0.156$ while the critical value is $\epsilon_c = 0.128$ (shown by the dashed black line).
Here $h = 0.3$, $\tau = 0.6$ and $\mathcal{T} = 100$.
}}
\label{fig mag_scaling}
\end{figure} 

\section{Order Parameter}
\label{secIV}
To investigate the DTC phase, we examine the stroboscopic dynamics of the magnetization along the $z$-axis, defined as
\begin{align}
m_z(n, \epsilon) = \frac{1}{N} \langle \psi_n(\epsilon) | S_z | \psi_n(\epsilon) \rangle,
\end{align}
where $n$ denotes the stroboscopic time (number of Floquet cycles). Fig.~\ref{fig mag}(a) presents the evolution of $m_z(n, \epsilon)$ for last $40$ periods for different values of $\epsilon$. For small $\epsilon$, the magnetization exhibits persistent oscillations with period $2\tau$, signaling robust DTC order. As $\epsilon$ increases, these oscillations gradually decay, and the system transitions to a non-oscillatory regime. 
The presence of period doubling is confirmed by the pronounced peak at frequency $\omega = \pi/\tau$ in the subharmonic Fourier spectrum, as illustrated in the inset to Fig.~\ref{fig mag}(a). Fig.~\ref{fig mag}(b) displays $m_z(n, \epsilon)$ as a function of $\epsilon$ for several time steps $n = 200$ to $221$, revealing a pronounced suppression of oscillations as $\epsilon$ increases and indicating a transition from DTC to trivial (non-DTC) behavior. To locate this boundary more precisely, we compute the susceptibility,
\begin{align}
\chi(n, \epsilon) = \frac{\partial}{\partial \epsilon} m_z(n, \epsilon) . 
\end{align}
The inset of Fig.~\ref{fig mag}(b) presents both the stroboscopic magnetization and the susceptibility as a function of  $\epsilon$, fixing $n=101$ and $N=300$. A sharp dip in susceptibility is evident at $\epsilon_c \approx 0.128$, suggesting a second-order phase transition dividing the DTC and non-DTC regimes.

For a quantitative characterization of the nature of the transition, we define an order parameter as the long-time average of the revived magnetization:
\begin{align} \label{eqn order param}
\overline{m}_z(\epsilon) = \frac{1}{\mathcal{T}} \sum_{n=0}^{\mathcal{T}} m_z(2n, \epsilon),
\end{align}
where $\mathcal{T}$ is the total number of observed stroboscopic pulses. Fig.~\ref{fig mag_scaling}(a) depicts $\overline{m}_z(\epsilon)$ as a function of $\epsilon$ for various system sizes $N$.  For $\epsilon < \epsilon_c$, the order parameter remains finite, indicating the presence of long-lived subharmonic response and DTC order. For $\epsilon > \epsilon_c$, the order parameter vanishes, reflecting the destruction of time-crystalline order.  This behavior, together with the sharpening of the transition for growing $N$, suggests a continuous (second-order) phase transition in the thermodynamic limit.

In the thermodynamic limit, close to a continuous transition, the order parameter is expected to vanish algebraically as
\begin{align} \label{eq: critical exponent}
\overline{m}_z(\epsilon) \sim |\epsilon - \epsilon_c|^{-\zeta_m},
\end{align}
where $\zeta_m$ is a critical exponent. This scaling is governed by the divergence of a correlation length $\xi \sim |\epsilon - \epsilon_c|^{-\nu_m}$ with $\nu_m$ as another critical exponent. For finite-size systems, the order parameter acquires system-size-dependent corrections, resulting in the finite-size scaling form
\begin{align}
\overline{m}_z(\epsilon) = N^{\zeta_m/\nu_m} f\left[N^{1/\nu_m} (\epsilon - \epsilon_c)\right],
\label{mzscaling}
\end{align}
where $f[\cdot]$ is a universal scaling function. To extract the critical exponents $\zeta_m$ and $\nu_m$, we perform the optimal collapse of $\overline{m}_z(\epsilon)$ over system sizes $N=60$ to $1000$, as shown in Fig.~\ref{fig mag_scaling}(b). We utilize finite-size scaling techniques such as \texttt{jaxfss}~\cite{yoneda2023_neural} (or \texttt{pyfssa}~\cite{sorge2015_pyfssa}), optimizing for the best collapse to extract the critical parameters: $ \epsilon_c = 0.128, \, \nu_m = 2.369, \, \zeta_m = -0.156$. and confirm that the DTC-to-trivial transition is continuous. Furthermore, the successful scaling collapse validates $\overline{m}_z(\epsilon)$ as an effective order parameter, situating the observed transition within the broader context of equilibrium-like critical phenomena in periodically driven systems.  We compare the above results with those obtained through mean-field analysis in the thermodynamic limit in Appendix. \ref{app:classical}, and also verify the value of $\epsilon_c$ through the analysis of Floquet states in Appendix \ref{appFloquet}.

\section{DTC Sensing through QFI}
\label{secV}

The continuous (second-order) nature of the DTC to non-DTC transition makes the system an attractive platform for quantum sensing: close to a critical point, any small perturbation of the control parameter is encoded in macroscopically large variations of the state, leading to an enhanced QFI. In what follows, we treat the spin-flip imperfection~$\epsilon$ as the parameter to be estimated.

\begin{widetext}
\begin{figure}[H]
\centering
\includegraphics[width=\textwidth]{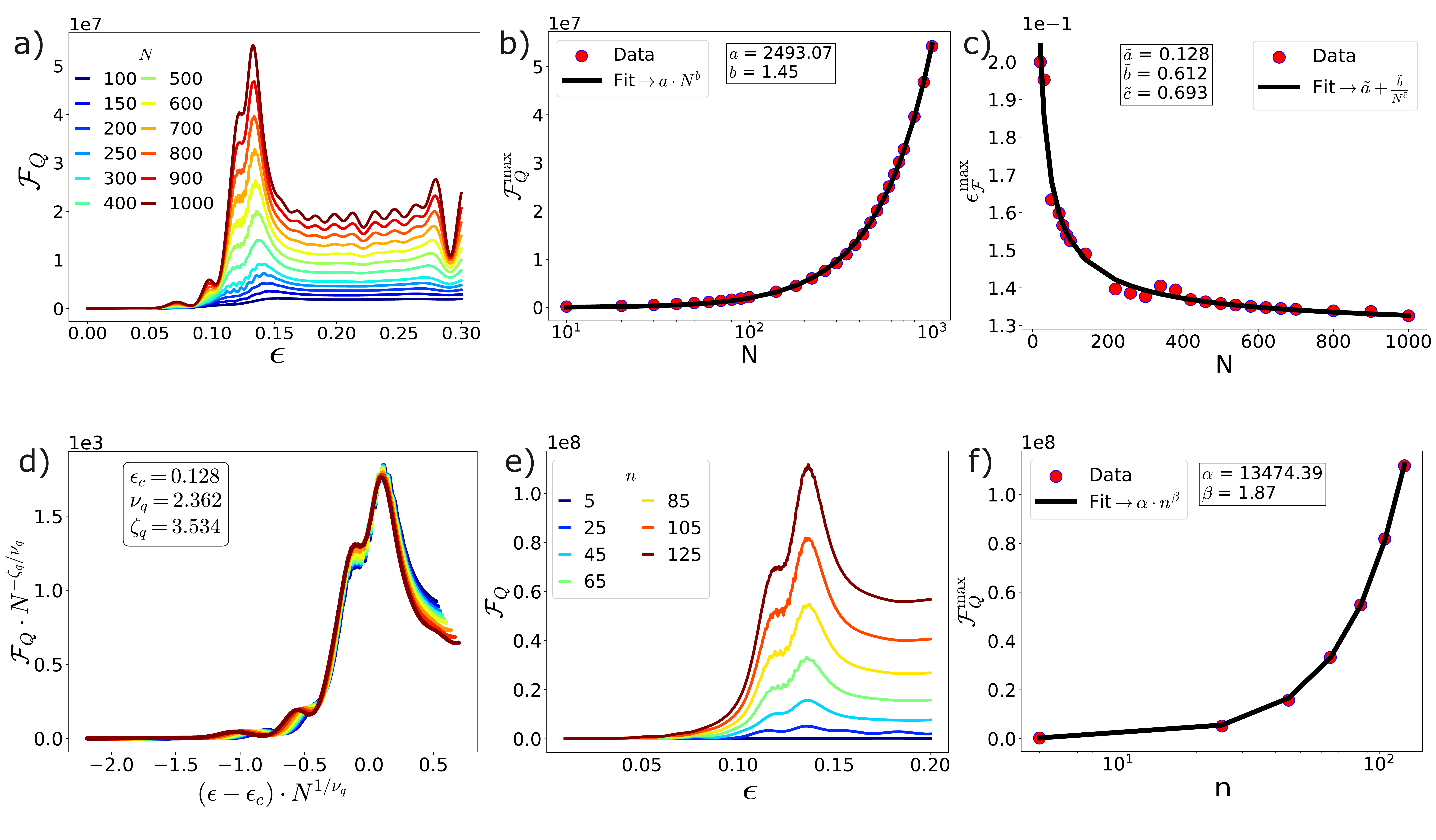}
 \begin{minipage}{0.999\textwidth}
\caption{ \justifying{{\bf Quantum Fisher Information:}
(a) QFI $\mathcal{F}_Q$  as a function of $\epsilon$ for various system sizes $N$ at $n=50$.
(b) The maximum QFI $\mathcal{F}_Q^{\text{max}} = \mathcal{F}_Q (\epsilon^{\text{max}}_{\mathcal{F}})$ as a function of $N$. We fit the data to a function of the form $ \mathcal{F}_Q = a \cdot N^b $ with $a=2493.07$, and $b=1.45$.   
(c) The value of $\epsilon^{\text{max}}_{\mathcal{F}} $  as a function of $N$, fitted to a Pareto-like function $ \epsilon^{\text{max}}_{\mathcal{F}}  = ( \tilde{a}  + \frac{\tilde{b}}{N^{\tilde{c}}} )$ with $\tilde{a}=0.128$, $\tilde{b}=0.612$, and $\tilde{c} = 0.693$. 
(d) Finite-size scaling analysis that reveals the best data collapse using critical parameters $(\eps_c = 0.128, \nu_q = 2.362, \zeta_q = 3.534)$. 
(e) Dynamic growth of the QFI $\mathcal{F}_Q $ versus $\eps$ for different $n$ in a system of size $N=500$. 
(f) QFI $\mathcal{F}_Q^{\rm{max}} $ as a function of time steps $n$. We fit the data to a function of the form $ \mathcal{F}_Q^{\rm{max}} = \alpha \cdot n^{\beta} $ with $\alpha=13474.39$, and $\beta = 1.87$. 
We have set, for all, $h=0.3$, and $\tau = 0.6$. 
} }
\label{fig QFI}
\end{minipage}
\end{figure}  
\end{widetext}

Fig.~\ref{fig QFI}(a) shows the numerically evaluated QFI $\mathcal{F}_Q $ after $n=50$ Floquet cycles for system sizes $N=100,\dots,1000$. For all $N$, the QFI grows with~$\epsilon$ till it reaches a pronounced maximum at the transition point $\epsilon = \epsilon^{\text{max}}_{\mathcal{F}}$ with the value $\mathcal{F}_{Q}^{\text{max}}$, and then it falls off rapidly. Furthermore, the entire QFI profile gets enhanced as the system size $N$ (and likewise the stroboscopic time $n$, shown below) is increased, anticipating genuine quantum-metrological scaling that we quantify below. 
We have plotted in Fig.~\ref{fig QFI}(b) the maximum QFI, $\mathcal{F}_Q^{\text{max}}$, as a function of $N$ and have fitted them with a function $\mathcal{F}_{Q}^{\text{max}} \approx a \cdot N^b$. The fitting parameters are $ a = 2493.07 $, and $ b = 1.45 $.  The value $b > 1$ demonstrates \emph{quantum-enhanced sensitivity}, i.e., super-linear scaling that surpasses the SQL, as $\mathcal{F}_{Q}^{\text{max}}$ diverges with the same exponent as we increase the system size. To see how the transition point $\epsilon^{\text{max}}_{\mathcal{F}}$ varies with $N$ as well as to estimate its value in the thermodynamic limit ($N \to \infty$), we have plotted the same as a function of the system size $N$ in Fig.~\ref{fig QFI}(c). We can precisely map the data to a fitting function,  
\begin{align}
\epsilon^{\text{max}}_{\mathcal{F}} \approx \tilde{a} + \frac{\tilde{b}}{N^{\tilde{c}}}
\label{Fscaling1}
\end{align}
where $\tilde{a}$ represents the critical imperfection in the thermodynamic limit i.e $ \tilde{a} =  \epsilon^{\text{max}}_{\mathcal{F}} (N = \infty) = \epsilon_c $, and $\tilde{c}$ depicts how fast the second part of the fitting function is going to vanish with increasing system size while $\tilde{b}$ is a proportionality constant. Our analysis yields the parameters $\tilde{a} = 0.128$, $\tilde{b} = 0.612$, and $\tilde{c} = 0.693$.

To study the observed transition, we start with an ansatz for the  QFI fixing $n$ in the following form \cite{montenegro24quantum}
\begin{align} \label{eqn qfi_fss_1}
\mathcal{F}_Q = \frac{a }{ N^{-b} + c \left(\epsilon - \epsilon^{\text{max}}_{\mathcal{F}} \right)^{\zeta_q} }, 
\end{align}
with some constant $a$, $b$, $c$, and $\zeta_q$, which need to be evaluated. The second-order character of the transition implies that all relevant quantities, including the QFI, are expected to exhibit scale invariance in the vicinity of the critical point. At $\epsilon = \epsilon^{\text{max}}_{\mathcal{F}}$, one can retrieve $\mathcal{F}^{\text{max}}_Q \approx a \cdot N^b$. On the other hand, for $N \to \infty$, one can obtain $\mathcal{F}_{\infty} \propto |\epsilon - \epsilon^{\text{max}}_{\mathcal{F}}|^{-\zeta_q}$. To estimate $\zeta_q$, we perform a finite-size scaling analysis. So, we take the standard scaling form
\begin{align} \label{eqn qfi_fss_2}
\mathcal{F}_Q  = N^{\frac{\zeta_q}{\nu_q}} f\left( N^{\frac{1}{\nu_q}} \ (\epsilon - \epsilon_c) \right)   .
\end{align}

In Fig.~\ref{fig QFI}(d), we plot $\mathcal{F}_Q \cdot N^{-\frac{\zeta_q}{\nu_q}}$ as a function of $N^{\frac{1}{\nu_q}} \cdot (\epsilon - \epsilon_c)$ for various system sizes. Using the numerical techniques mentioned above, we determine the critical point $\epsilon_c = 0.128$, the critical exponents $\nu_q = 2.362$ and $\zeta_q = 3.534$. At $\epsilon = \epsilon^{\text{max}}_{\mathcal{F}} = \epsilon_c$, one can obtain $\frac{\zeta_q}{\nu_q} = b$ by comparing equations \eqref{eqn qfi_fss_1} and \eqref{eqn qfi_fss_2} \cite{montenegro23quantum}. From the two different analyses depicted above, we have $b = 1.45$, which agrees closely with $\frac{\zeta_q}{\nu_q} = 1.496$. 

Following Ref.~\cite{yoneda2023_neural}, to study the dependence of the QFI on the time steps $n$, we plot the QFI as a function of $\epsilon$ for various values of $n$ in Fig.~\ref{fig QFI}(e), fixing $N=500$. The same nature of the QFI curve, which we have seen with $N$, is present here for $n$. It is evident that the QFI increases with $n$ for all values of $\epsilon$; however, near the transition point $\epsilon_c$, the scaling of QFI with $n$ reaches its maximum value. To check this scaling with $n$, we plot the QFI corresponds to $\epsilon^{\text{max}}_{\mathcal{F}}$, and map the data with the function $ \mathcal{F}_Q^{\text{max}} \propto n^{\beta}$.  Finally, based on the above analysis, one can write the following ansatz for the QFI 
\begin{align}
\mathcal{F}_Q (\epsilon, n, N)  = N^{\frac{\zeta_q}{\nu_q}} f\left( N^{\frac{1}{\nu_q}} \ (\epsilon - \epsilon_c) \right) \, n^{\beta} 
\end{align}
with $b = \frac{\zeta_q}{\nu_q}  \simeq 1.45$ and numerical fitting gives $\beta \simeq 1.87$.  It is worth emphasizing that classical probes achieve at best $b = 1$ and the Heisenberg limit corresponding to $b = 2$ \cite{montenegro24quantum, biswas2025_floquet}. We highlight this as one of the main results of this article, demonstrating that our DTC probe achieves quantum-enhanced sensitivity. 

\section{Melting transition of the DTC through inverse participation ratio}
\label{secVI}
The melting of the DTC phase at a critical value, observed above, was first reported in Ref.~\cite{russomanno17floquet}. More recently, Ref.~\cite{munoz_arias_2022_floquet} presented a comprehensive study of the Floquet dynamics in periodically driven collective $p$-spin models. This investigation employed higher-order correlation functions-specifically, infinite time-averaged out-of-time-order correlators-to characterize the transition between non-Floquet time crystal (non-DTC) and Floquet time crystal (DTC) phases. In an alternative approach, Liu et al.~\cite{liu2023_discrete} utilized degenerate perturbation theory to analyze the overlap between initial product states and the quasi-eigenstates of the Floquet operator. Their findings suggest that the total height of the peak in the Fourier spectrum is determined by the eigenspace inverse participation ratio. Motivated by these findings, here we employ the time-averaged inverse participation ratio (IPR) as a diagnostic tool for probing the melting of the DTC phase with increasing imperfection strength. In dynamical systems, the IPR offers a more sensitive measure of localization effects~\cite{lezama2021_equilibration, jacob2019_manybody, nicolas2019_maltifractal, liu2023_probing}. IPR $\mathcal{I}(n, \epsilon)$, as a nonlocal quantity, evaluated in a chosen preferential basis, is formally defined as~\cite{lezama2021_equilibration, liu2023_probing}
\begin{align}
\mathcal{I}(n, \epsilon) = \sum_{i} | \langle \mathbf{z}_i | \psi_n \rangle|^4
\label{eqlocalization}
\end{align}
where $| \psi_n \rangle $ is the normalized  $n$-th stroboscopic state which is defined in Eq. \eqref{eqpsin}, and the $| \mathbf{z}_i \rangle$ denotes the computational basis. The latter quantifies the spreading of the initial state $|\psi_0 \rangle$ across the many-body Hilbert space after the $n$-th stroboscopic time. If the $n$-th stroboscopic state coincides with a single basis state in the set $\{ | \mathbf{z}_i \rangle \}$, the IPR will be the maximum ($= 1$), indicating a localized state. On the other hand, for a state represented by a linear combination of the basis vectors, the IPR is less than one, and approaches the limit of $\frac{1}{\mathcal{D}}$ for a maximally randomized state in the basis set,  where $\mathcal{D}$ is the dimension of the associated Hilbert space.

\begin{figure}
\centering
\includegraphics[width=\linewidth]{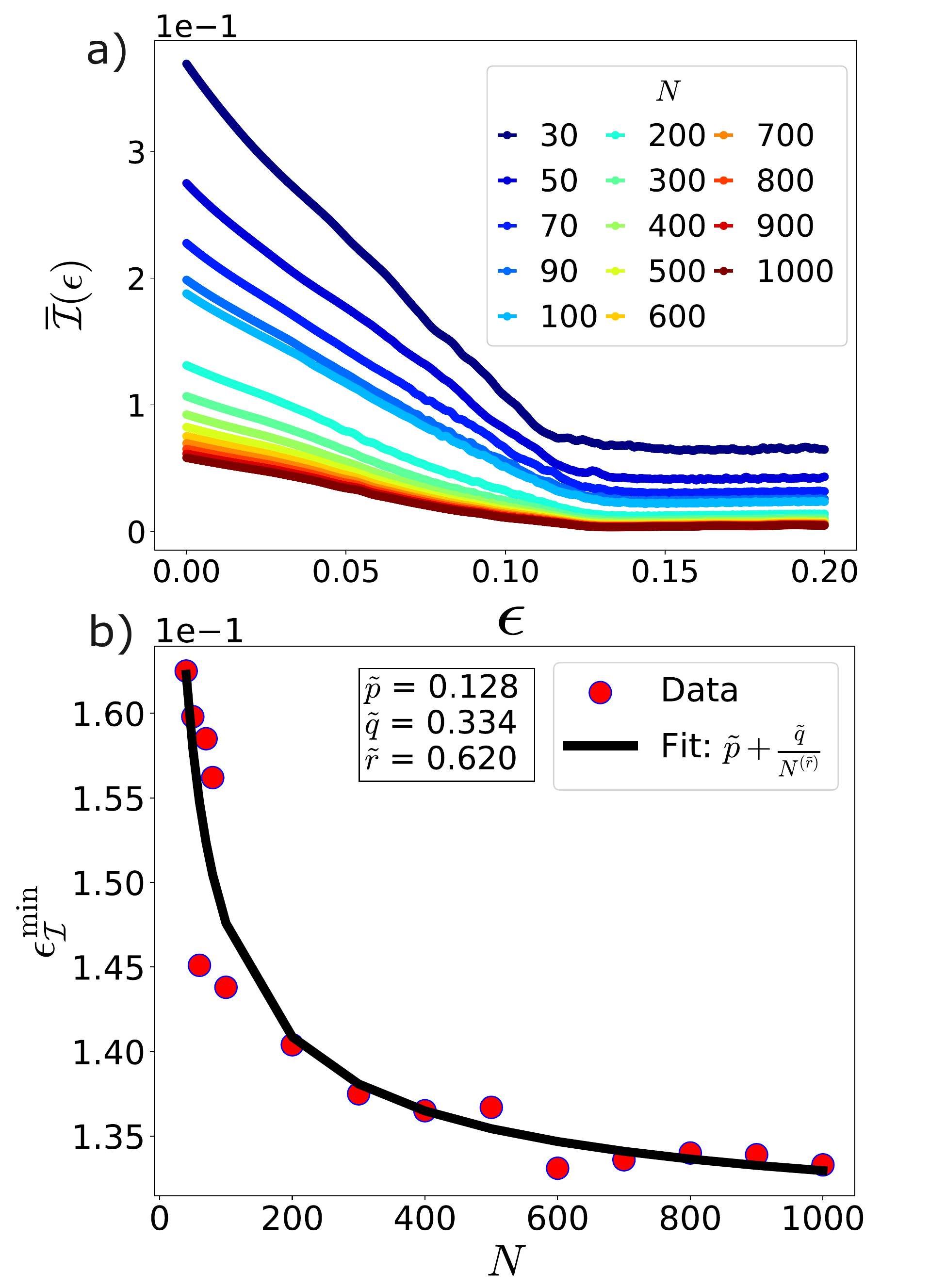}
    \caption{ \justifying{{\bf Time-averaged inverse participation ratio:}
 (a) TAIPR $\overline{\mathcal{I}} (\epsilon)$  as a function of $\epsilon$ for various system sizes $N$. 
 (b) $\eps^{\text{min}}_{\mathcal{I}}$  as a function of $N$, fitted to a Pareto-like function $ \eps^{\text{min}}_{\mathcal{I}} (N) = ( 0.128 + \frac{0.334}{N^{0.62}} )$. Here $h=0.3$, $\tau = 0.6$, and $\mathcal{T} = 100$.}}
    \label{fig IPR}
\end{figure}

To understand this transition property, we define a time-averaged IPR (TAIPR): 
\begin{align}
\overline{\mathcal{I}} (\epsilon) =  \frac{1}{\mathcal{T}} \sum_{n=0}^{\mathcal{T}} \mathcal{I}(n, \epsilon),
\label{eqTaipr}
\end{align}
which quantifies the  spreading of the initial state. Here $\mathcal{T}$ represents the total number of time steps. In Fig.~\ref{fig IPR}(a), we present $\overline{\mathcal{I}}$ as a function of $\epsilon$. The results demonstrate that the TAIPR is large for $\epsilon < \epsilon_c$, signaling the localization of the time evolved state into a few quantum states (see Eqs. \eqref{eqlocalization} and \eqref{eqTaipr}). On the other hand,  as  $\epsilon$ increases, the TAIPR decreases sharply  till  $\epsilon = \epsilon^{\text{min}}_{\mathcal{I}}$, beyond which $\overline{\mathcal{I}}$ settles into an approximately $\epsilon$-independent value $\overline{\mathcal{I}}^{\text{min}}$. This behavior indicates that beyond $\epsilon^{\text{min}}_{\mathcal{I}}$, the initial state becomes maximally spread across the many-body Hilbert space within the considered time frame, which allows us to identify $\epsilon^{\text{min}}_{\mathcal{I}}$ with $\epsilon_c$. 
 In Fig.~\ref{fig IPR}(b), we plot the values  $\epsilon^{\text{min}}_{\mathcal{I}}$ as a function of system size $N$ and map to a fitting function
\begin{align}
\epsilon^\text{min}_{\mathcal{I}} = \tilde{p} + \frac{\tilde{q}}{N^{\tilde{r}}}  ,
\end{align}
with parameters $ \tilde{p}  = 0.128$, $\tilde{q} = 0.334$, and $\tilde{r} = 0.62$. The parameter $\tilde{p}$ denotes the critical value of the imperfection at thermodynamic limit i.e. $  \tilde{p} =  \epsilon^\text{min}_{\mathcal{I}} (N \to \infty) = \epsilon_c $ while the second part becomes zero, therefore once more highlighting $\epsilon \approx 0.128$ as the critical point.

\begin{widetext}
\begin{figure}[htbp]
\centering
\includegraphics[width=\textwidth]{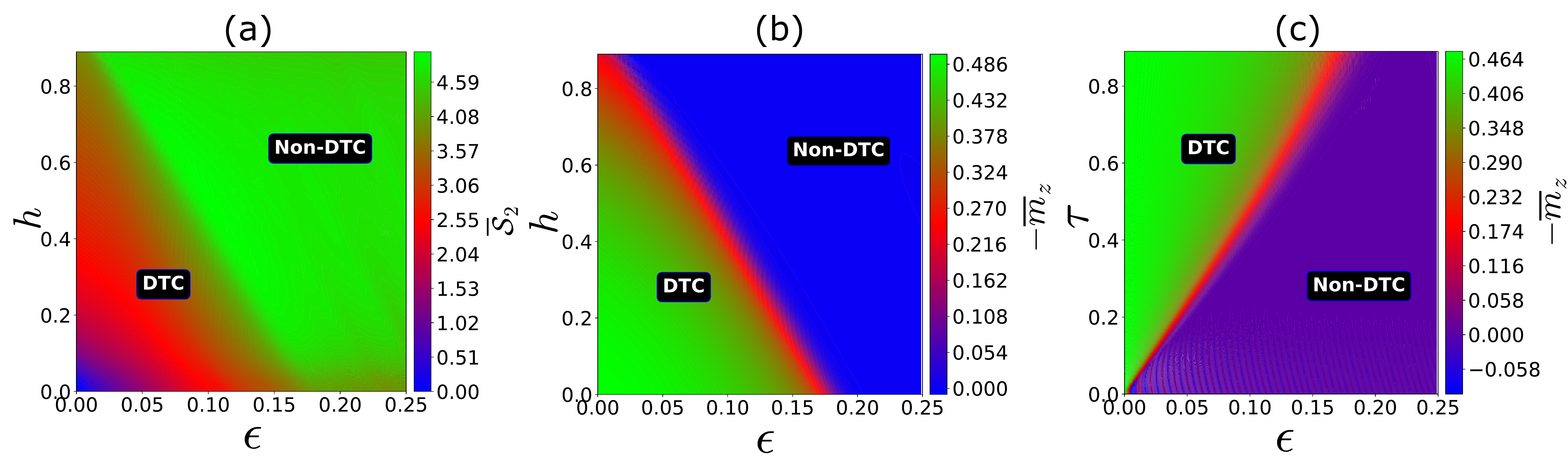}
\begin{minipage}{0.999\textwidth}
\caption{  \justifying{ {\bf Dynamical phase diagrams:}
(a) The second-order Rényi entropy, $\overline{\mathcal{S}}_2$, as a function of magnetic field $h$ and spin-flip imperfection $\epsilon$ with $\tau=0.6$. 
(b) Negative of the order parameter, $-\overline{m}_z$, as a function of  $h$ and  $\epsilon$ with $\tau=0.6$. 
(c) Negative of the order parameter, $-\overline{m}_z$, as a function of period of the drive $\tau$ and $\epsilon$ fixing $h=0.3$.
The other parameters are $N=400$, and $n=100$. 
} }
\label{fig phase diagram}
\end{minipage}
\end{figure}

\end{widetext}

\section{Effect of Magnetic field and period}
\label{secVII}
In this section, we investigate how the magnetic field, $h$ influences the melting of the DTC phase. Building on the previous analysis, we use the TAIPR as a reliable probe to locate the critical spin-flip imperfection and extend this approach to map the transition in the $(\epsilon, h)$ plane. Fig.~\ref{fig phase diagram}(a) shows a density plot of the time-averaged second-order Rényi entropy \cite{atas2012_multifractality, luitz2017_ergodic, laflorencie2016_quantum}, 
\begin{align}
   \overline{\mathcal{S}}_{2} = -\ln (\overline{\mathcal{I}}) ,
\end{align}
as a function of $\epsilon$ and $h$. The diagram reveals that the critical imperfection $\epsilon_c$ systematically decreases with increasing $h$, finally vanishing close to the $\mathbb{Z}_2$ symmetry breaking transition at $h = J = 1$.  This indicates that stronger quantum fluctuations arising from non-commuting $S_x$ and $S_z$ parts in $H_1$ accelerate the delocalization of the initial state in the many-body Hilbert space in finite-size systems, thereby accelerating the destruction of the DTC phase ((see Eq. \eqref{eq: floquet hamiltonian} and Appendix \ref{AppB})).

For comparison, Fig.~\ref{fig phase diagram}(b) presents the corresponding order parameter $\overline{m}_z$ over the same parameter range. The transition line extracted from $\overline{m}_z$ coincides with that obtained from the time-averaged second-order Rényi entropy analysis, confirming the consistency of the two diagnostics.

For the sake of completeness, we present the order parameter $\overline{m}_z$ as a function of the period of the drive, $\tau$ and $\epsilon$. It appears that the higher value of $\tau$ widens the DTC region. As $\tau$ increases, the spins interact  for a longer time (see Eq. \eqref{eq: floquet hamiltonian}), which results in the DTC phase being robust to  higher values of imperfection.

\begin{widetext}
\centering
\begin{tabular}{|c|c|c|c|}
\hline
Quantity & Scaling forms & Parameters Values &  Transition Points   \\
  &   &   &   from FSSA   \\
\hline
Order Parameter, $\overline{m}_z (\epsilon) $ & $ \overline{m}_z  = N^{\frac{\zeta_m}{\nu_m}} f\left( N^{\frac{1}{\nu_m}} \ (\epsilon - \epsilon_c) \right) $ & $ \nu_m = 2.369,  \zeta_m = -0.156$ &  $ \epsilon_c =  0.128 $ \\
\hline
QFI, $\mathcal{F}_Q (n, \epsilon) $ & $\mathcal{F}_Q  = N^{\frac{\zeta_q}{\nu_q}} f\left( N^{\frac{1}{\nu_q}} \ (\epsilon - \epsilon_c) \right) \, n^{\beta} $ & $\nu_q = 2.362, \zeta_q = 3.534, \beta = 1.87$ &  $ \epsilon_c =  0.128$ \\
\hline
\end{tabular}
\captionof{table}{Scaling behavior and critical parameters for various quantities near the quantum phase transition. Here, FSSA stands for finite-size scaling analysis, and the numerical analysis is done for $h = 0.3$ and $\tau = 0.6$.}
\label{tab:scaling_parameters}
\end{widetext}

\section{Conclusion}
\label{secVIII}

In this paper we have studied the critical properties of DTC to non-DTC transition in the LMG model subjected to a periodic modulation in time, and then used this transition to model high-precision quantum sensors. Notably, we show that the QFI exhibits superlinear scaling with the system size $N$, thereby beating the SQL, and allowing the possibility of engineering high-performing quantum sensors  close to the DTC phase transition. Crucially, the metrological advantage is not merely a consequence of the static criticality at
$h = J$, but stems directly from the phase transition at $\epsilon_c$ with $h < J$, signaling the destruction
of the time-crystalline phase. Importantly, as shown in Figs. \ref{fig phase diagram}a, \ref{fig phase diagram}b and \ref{fig:classical_transition}b, $\epsilon_c \to 0$ as we
approach the LMG criticality at $h = J$, thus implying that being in the vicinity of the  non-driven LMG
critical point is detrimental to the existence of a DTC phase with finite $\epsilon$. On the other hand,
 $h/J$ values smaller than unity, or equivalently, stronger inter-spin interaction strengths, are
associated with larger values of  $\epsilon_c$.

We have studied the dynamics of the system with increasing $\epsilon$. As expected, the DTC phase exists only for smaller values of $\epsilon$.  Our detailed analysis of order parameter $\overline{m}_z$, the corresponding susceptibility, QFI and TAIPR suggest a critical point at $\epsilon \approx 0.128$ for the system parameters considered here, beyond which the DTC phase is lost. Furthermore, we have done finite-size scaling analysis of $\overline{m}_z$ and QFI, and evaluated the corresponding critical exponents.  We have summarized the results in Table \ref{tab:scaling_parameters}.  The continuously changing form of $\overline{m}_z$ close to $\epsilon = \epsilon_c$ (see Figs. \ref{fig mag_scaling} and \ref{fig:classical_transition}), closely lying values of the critical point $\epsilon_c$ and the correlation length exponent $\nu_m \approx \nu_q$, and the agreement of the finite-size scaling analysis  results with those obtained in the thermodynamic limit (see Appendix \ref{app:classical}, Figs. \ref{fig:classical_transition} and \ref{fig: poincare maps}) emphasize the second order nature of this transition, and the consistency of our analysis.  Finally, we have studied the dynamical phase diagram of the system as functions of $\epsilon$, the transverse field $h$ and the time period $\tau$, using the magnetization and the Rényi entropy. Notably, the critical point shifts with changing $h$ and $\tau$.  Furthermore,  as confirmed by our analysis, the DTC phase exists only in the $\mathbb{Z}_2$ symmetry broken phase $h < J$, with the ideal $\pi$-kick exchanging the two symmetry-broken states;  higher magnitudes of quantum fluctuations arising for larger values of $h ~(< J = 1)$, and low values of the modulation time period $\tau$, are detrimental to the formation of the  DTC phase.

We emphasize that the value $\epsilon_c \approx 0.128$ reported in the present work characterizes the loss of robust period-doubled response for the chosen initial state. As discussed in Appendix \ref{app:classical} and also in Ref. \cite{russomanno17floquet}, analysis of the phase space plot in the thermodynamic limit shows the existence of symmetry broken curves for small $\epsilon$; initialization of the system in one such symmetry broken curve, along with the periodically modulated Hamiltonian \eqref{eq: floquet hamiltonian}, results in the system traversing between these two symmetry broken curves with a period $2T$. On the other hand, the system ceases to lie on a symmetry broken curve for $\epsilon > \epsilon_c$, thereby destroying the DTC phase. However, as shown in Fig. \ref{fig: poincare maps}d, symmetry broken curves which are relevant for  other initial states may still exist for $\epsilon > 0.128$; in contrast, preliminary numerical analysis indicates the absence of any such symmetry broken curve  for larger values of $\epsilon$, irrespective of the choice of the initial state (see Fig. \ref{fig: poincare maps}e), thus suggesting the presence of an initial state independent criticality at large $\epsilon$. Relatedly, in addition to the non-DTC to $2$-DTC phase transition studied here, recent studies suggest
higher values of $\epsilon$ may correspond to additional higher order DTCs, which maybe prethermal in
nature \cite{mishra2025_coexistence, pizzi2021_higher}. Analysis of the possible critical properties of these higher-order DTC transitions and initial state independent criticality may pose interesting open questions. However, detailed studies of quantum sensing and critical exponents close to these additional higher order possibly prethermal DTC transitions is beyond the
scope of the present manuscript.

 Unlike many-body localization based time crystals \cite{yao17discrete}, the DTC studied here does not rely on disorder to prevent thermalization. Instead, here the long-range interactions play a crucial role in stabilizing the DTC \cite{russomanno17floquet, pizzi2021_higher}. Recent studies suggest long-range interactions in the presence of disorder can host a DTC phase in the presence of periodic driving \cite{bothra25discrete}. Consequently, studies of quantum sensing and scaling analysis close to DTC transition in long-range interacting systems in the presence of disorder can raise interesting questions as well.

We note that studies have suggested experimental realization of the LMG model using Bose-Einstein condensate in optical cavity setups \cite{chen09quantum}, or in cavity QED setups \cite{larson10circuit}. Furthermore, in the recent years DTCs have been experimentally realized in various setups, including nitrogen-vacancy centers in diamond \cite{choi2017observation}, ion traps \cite{zhang17a} and optical cavities \cite{taheri22all}. Therefore we expect the setup and dynamics considered here can be experimentally realized in the near future, using the currently existing platforms.

\section*{ACKNOWLEDGEMENTS}
R.G. is grateful to Sayan Choudhury and Atanu Rajak for insightful discussions. B.D. acknowledges support from Prime Minister Research Fellowship(PMRF). V.M. acknowledges support
from  Anusandhan National Research Foundation (ANRF) through ARG (Project No. ANRF/ARG/2025/002531/PS.

\appendix

\section{Derivative of function} 
\label{sec Appendix A}
For the evaluation of QFI (see Eq. \eqref{eqQFI}), we calculate the derivative of a wavefunction numerically using the fourth-order five-point method ~\cite{Qvarfort2018_gravimetry}. For a function $g(x)$ of a variable $ x $ and step-size $\delta x$,  the first derivative using this method is expressed as 
\begin{multline}
g^{\prime}(x) = \frac{1}{12h} \Big[ -g(x + 2\delta x) + 8 g(x + \delta x) \\  
- 8 g(x - \delta x) + g(x - 2\delta x) \Big] + O(\delta x^4).
\end{multline}
\begin{figure}[htbp]
\centering
\begin{subfigure}{0.9\linewidth}
\centering
\textbf{(a)}\\[-0.2em]
\includegraphics[width=\linewidth]{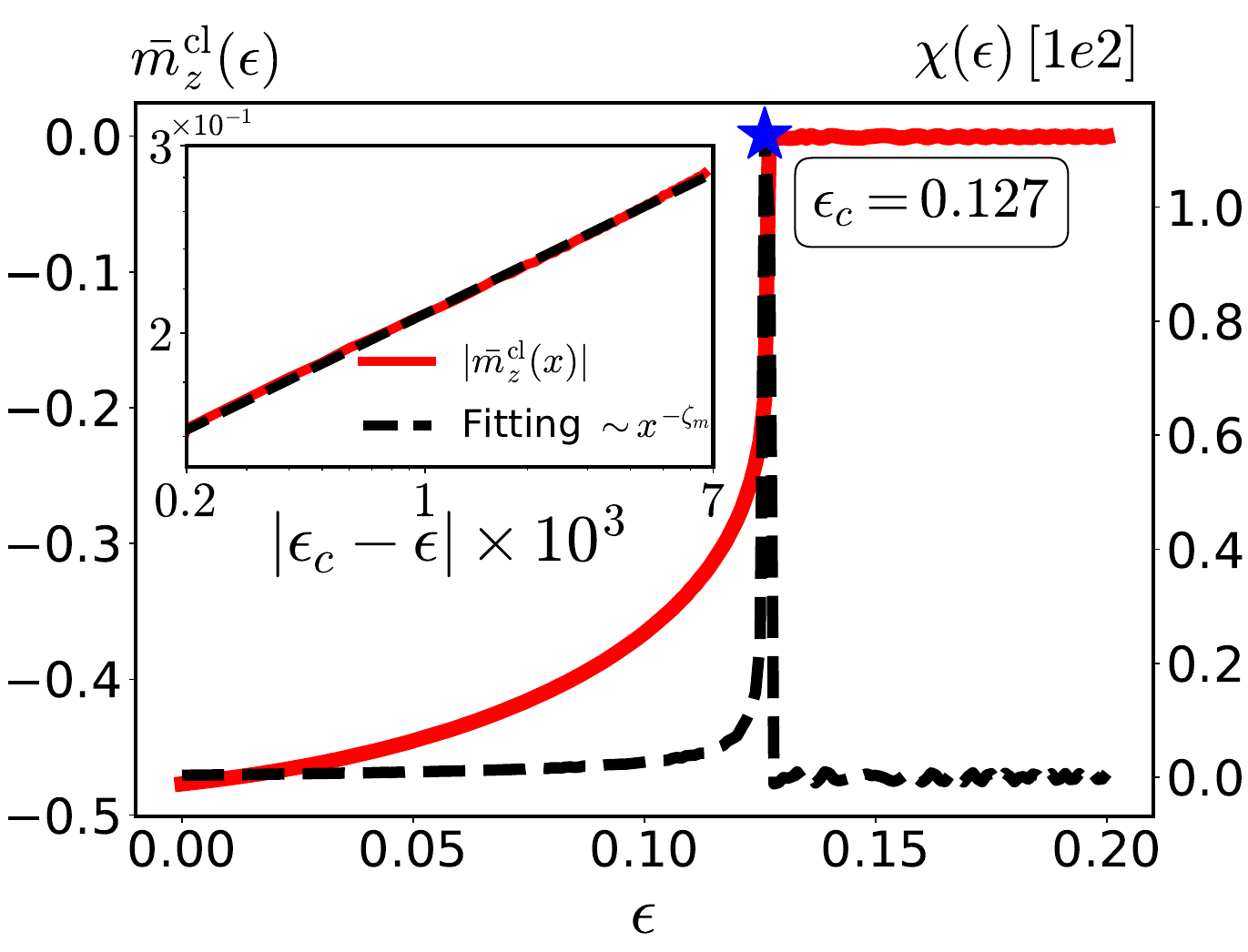}
\end{subfigure}
\begin{subfigure}{\linewidth}
\centering
\textbf{(b)}\\[0.2em]
\includegraphics[width=\linewidth]{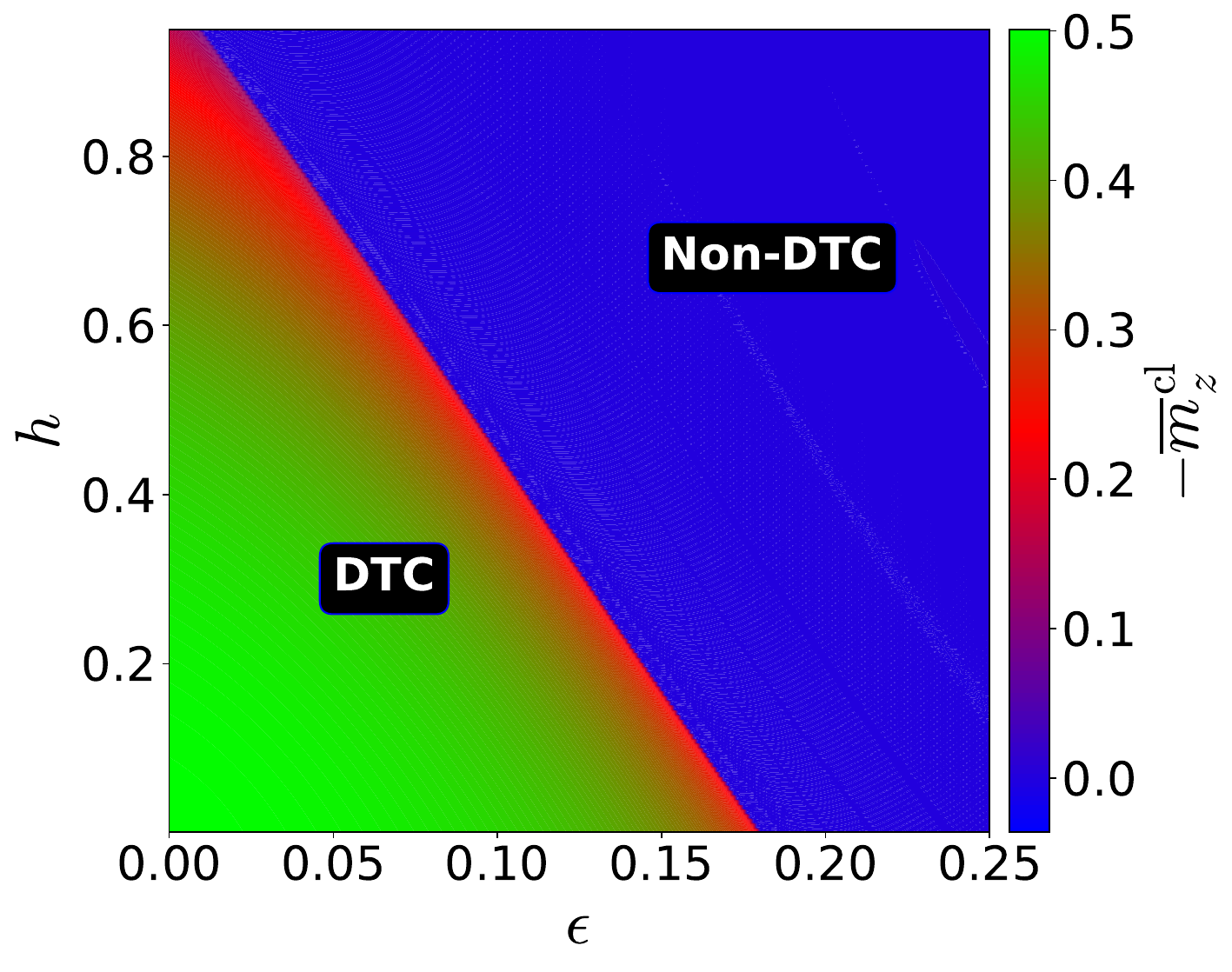}
\end{subfigure}

\caption{\justifying{{\bf (a)}: Classical (mean-field) order parameter $\bar{m}_z^{\rm cl}(n)$ (black dashed line) and classical susceptibility \(\bar{\chi}^{\rm cl}(n,\epsilon)\) (red solid line) with $n=101$, showing a sharp peak at \(\epsilon_c^{\rm cl}\simeq 0.127\) (blue star). {\bf (Inset:)} Log-Log plot of the absolute value of the classical order parameter $\bar{m}_z^{\rm cl}$ (red dashed line) is fitted using Eq.~\eqref{eq: critical exponent} with $\epsilon_c =0.1266$ and $\zeta_m = -0.156$. Parameters are $h = 0.3$, $\tau = 0.6$ and $\mathcal{T} = 1000$, same as in the quantum analysis.  {\bf (b)}: Density plot of the negative of the classical order parameter $-\overline{m}_z^{\rm{cl}} $ as a function of  $h$ and  $\epsilon$.}}
\label{fig:classical_transition}

\end{figure}

\section{Classical (mean-field) limit and comparison with the quantum transition}
\label{app:classical}

In this appendix we briefly discuss the classical (mean-field) limit of the periodically driven LMG model and compare it with the quantum results presented in the main text.
In the thermodynamic limit $N\to\infty$, the collective spin behaves as a classical angular momentum and the driven LMG dynamics reduces to a nonlinear Hamiltonian map on the Bloch sphere. Following Refs.~\cite{das2006_infinite, sciolla2011_dynamical, russomanno17floquet}, we parametrize the classical magnetization by canonical variables $(Q,P)$, with $Q\in[-1,1]$ and $P$ an angular variable, such that the components of the normalized magnetization $\mathbf{m}= \frac{1}{2}\sqrt{1-Q^2}\cos(2P), \frac{1}{2}\sqrt{1-Q^2}\sin(2P), \frac{1}{2}Q. $ The classical, periodically kicked Hamiltonian can be written as
\begin{equation}
H(Q,P,t)=H_{\text{LMG}}(Q,P)+H_2(Q,P)\sum_{n\in\mathbb{Z}}\delta(t-n\tau),
\label{eq:classical_H}
\end{equation}
with the time-independent part 
\begin{align}
H_{\text{LMG}}(Q,P) &= -\frac{J}{2}Q^2 - h\sqrt{1-Q^2}\cos(2P),\non\\
H_2(Q,P) &=  \frac{\phi}{2}\sqrt{1-Q^2}\cos(2P)
\label{eq:H0_Russomanno}
\end{align}
The $\mathbb{Z}_2$ symmetry of the quantum model corresponds to the invariance of $H_0$ under $(Q,P)\mapsto (-Q,-P)$. The second Hamiltonian induce an $x$-rotation (amplitude $\phi$).

We have followed the same protocol as for the finite-size case to find out the nature of the transition. Fig.~\ref{fig:classical_transition} shows the long-time
average of the revived magnetization, i.e. the order parameter \(\overline{m}_z^{\rm cl}(\epsilon)\) and \(\bar{\chi}^{\rm cl}(\epsilon)\) (= \(\frac{\partial \overline{m}_z^{\rm cl}(\epsilon)}{\partial \epsilon}\)) for the same driving parameters used in the finite-size calculations. A pronounced peak in \(\bar{\chi}^{\rm cl}\) identifies a classical critical imperfection \(\epsilon_c^{\rm cl}\approx 0.127\), beyond which the period-doubled response is lost and the trajectory approaches a non-oscillatory (non-DTC) regime. This classical threshold lies very close to the quantum critical value extracted through analysis of finite size systems in the main text (\(\epsilon_c\approx 0.128\)), indicating that the location of the melting transition is well-captured at the mean-field level, while finite-\(N\) quantum fluctuations primarily produce a small shift and rounding of the transition. We have also plotted the phase diagram which matches with the quantum analysis. Finally, we fit the classical order parameter $\bar{m}_z^{\rm cl}(\epsilon)$ using Eq.~\eqref{eq: critical exponent}, obtaining $\epsilon_c = 0.127$ and $\zeta_m = -0.156$. \emph{This serves as a final confirmation of the finite-size scaling analysis presented in the main text.}

\begin{widetext}

\begin{figure}[H]
\centering
\begin{panelTR}{0.19\linewidth}{a}
\includegraphics[width=\linewidth]{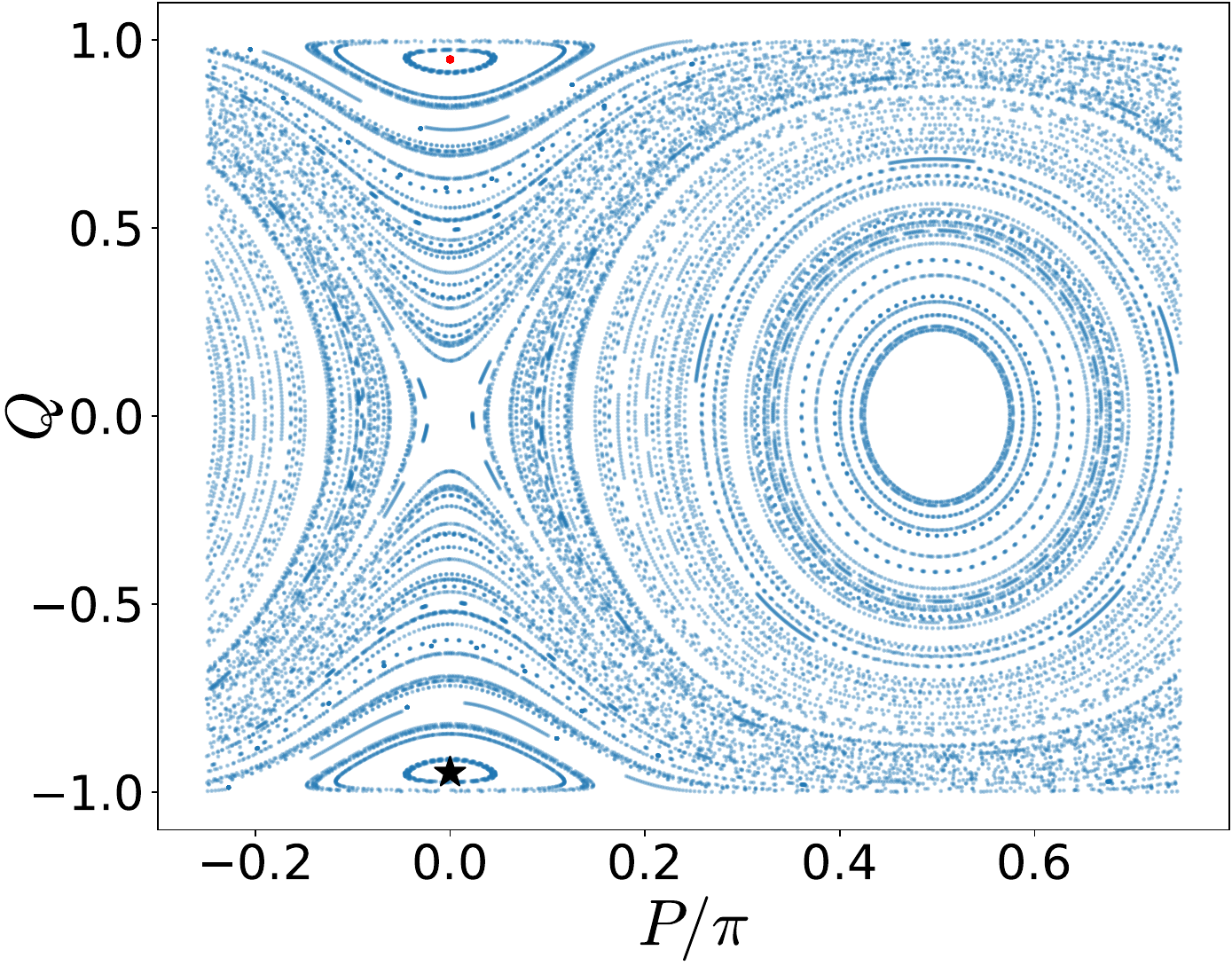}
\end{panelTR}
\hfill
\begin{panelTR}{0.19\linewidth}{b}
\includegraphics[width=\linewidth]{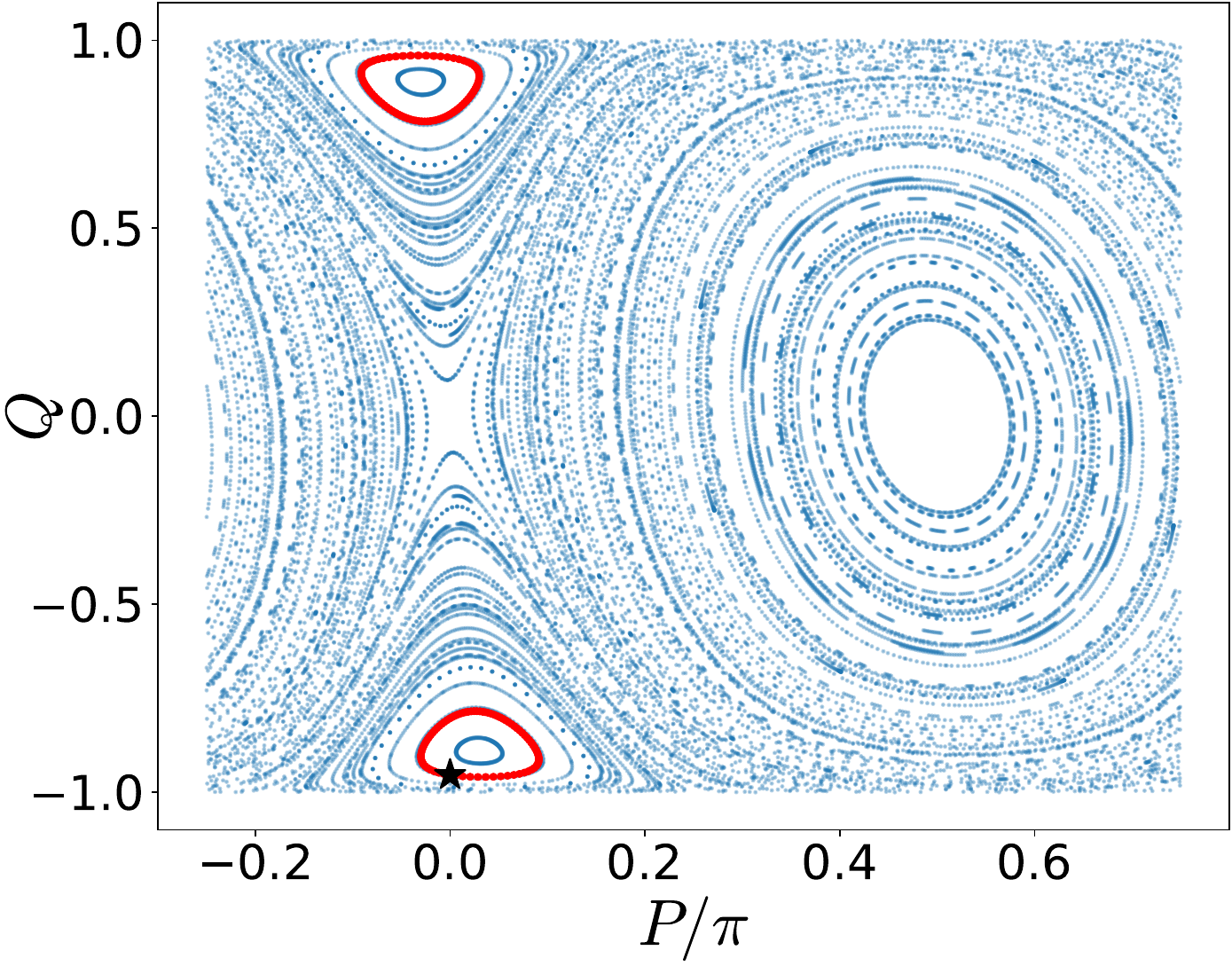}
\end{panelTR}
\hfill
\begin{panelTR}{0.19\linewidth}{c}
\includegraphics[width=\linewidth]{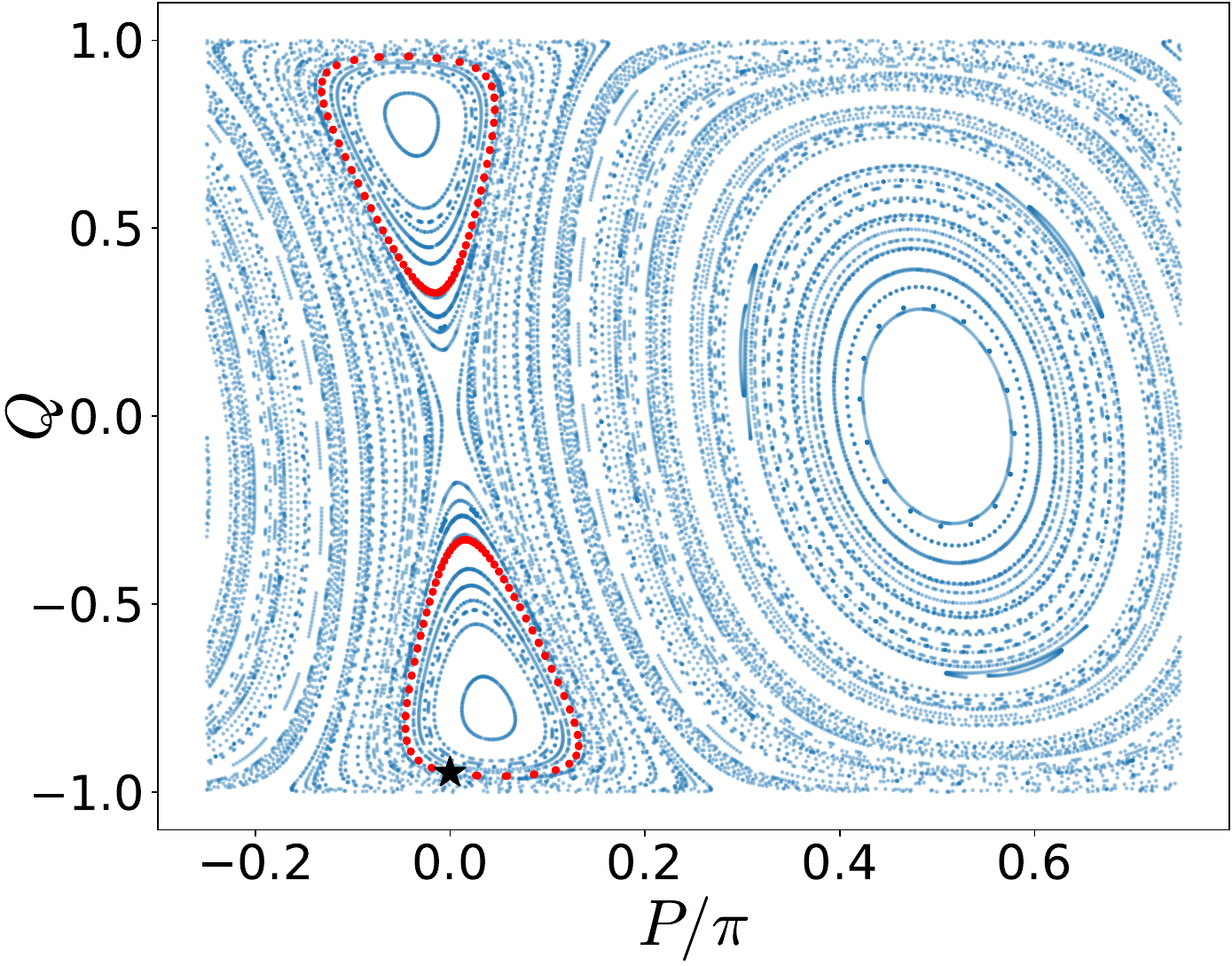}
\end{panelTR}
\hfill
\begin{panelTR}{0.19\linewidth}{d}
\includegraphics[width=\linewidth]{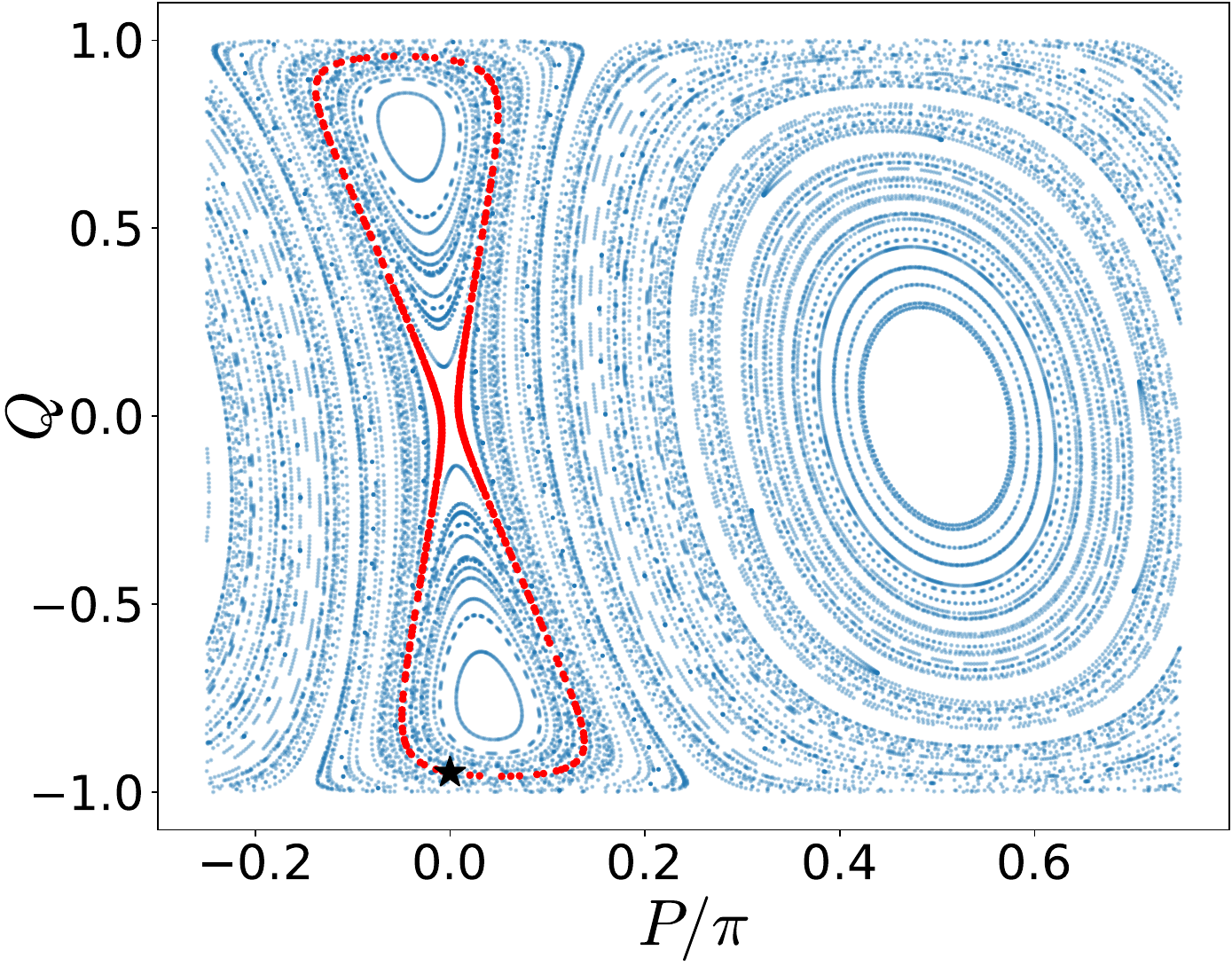}
\end{panelTR}
\hfill
\begin{panelTR}{0.19\linewidth}{e}
\includegraphics[width=\linewidth]{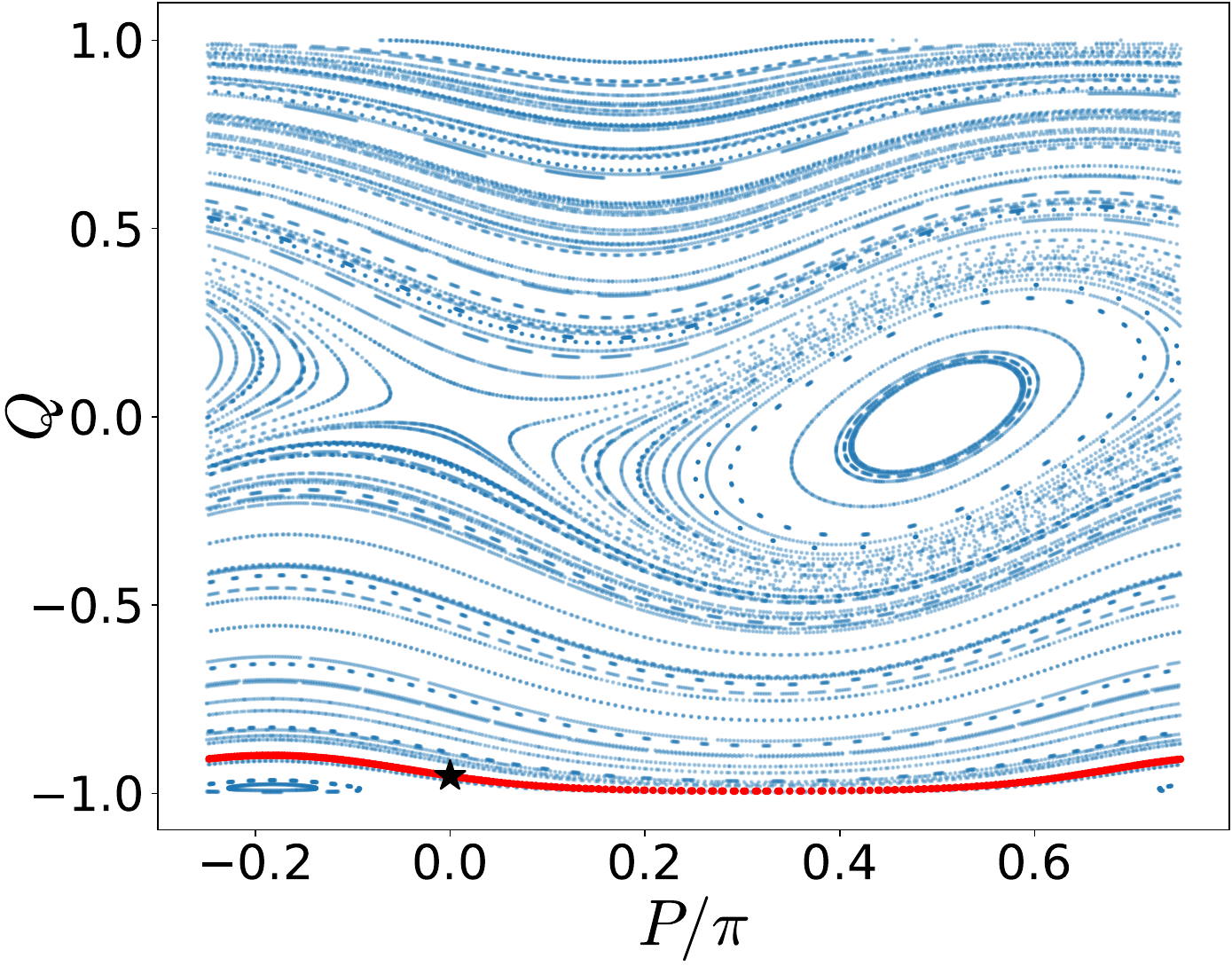}
\end{panelTR}
\caption{\justifying{Classical phase space for different values of the phase $\epsilon = 0, 0.06, 0.12, 0.13 {\color{blue}, 0.9}$ (from left to right). The black star point  denotes the initial symmetry-breaking ground state with $h$. The red circle points denote the stable islands around the period-doubled points which is vanishing around $\epsilon = 0.13$ for the chosen initial state.   As shown in (e), the period-doubling stable islands vanish completely for large values of $\epsilon$. Here $h=0.3$, and $\tau=0.6$.
}}
\label{fig: poincare maps}
\end{figure}
\end{widetext}

We show the phase space plot in Fig.~\ref{fig: poincare maps}. Consistent with observations in Ref. \cite{russomanno17floquet}, these plots reveal that the dynamics is regular, in the form of closed curves. Further, Fig. \ref{fig: poincare maps} shows the existence of curves that are symmetric under $P \to -P$, $Q \to -Q$, in addition to curves that break this symmetry. However, each symmetry-breaking curve has a symmetric partner at $P \to -P$, $Q \to -Q$. For small $\epsilon$, initialization of the system in one such symmetry-breaking curve, shown by the black star,  ensures that the system stays in that curve, till the periodic kick shifts it to its symmetric partner at $P \to -P$, $Q \to -Q$. On the other hand, the system lies on a curve that is invariant under $P \to -P$ and $Q \to -Q$ for $\epsilon = 0.13$, thus verifying the result $0.12 < \epsilon_c < 0.13$. Importantly, as shown in Fig. \ref{fig: poincare maps}e, all symmetry broken curves vanish  for larger values of $\epsilon$, thus resulting in non-period-doubling trajectories for all initial states in this regime. 

Additionally, we note that $H_{\rm LMG}$ (see Eq. \eqref{eq:H0_Russomanno}) has symmetry-broken minima at~\cite{das2006_infinite}
\begin{align}
P_0=0,\qquad Q_0=\pm\sqrt{1-(h/J)^2}.
\end{align}
Consequently, small oscillations about the minima of the form $Q \equiv Q_0 + \delta Q$ and $P = P_0 + \delta P$ gives rise to a harmonic oscillator Hamiltonian
\begin{align}
H_{\rm LMG}\simeq H_0+\frac12 K_Q(\delta Q)^2+\frac12 K_P(\delta P)^2,
\end{align}
for small $\delta Q, \delta P$, real $H_0$ and positive constants $K_{Q}$ and $K_P$ with $\sqrt{K_{Q} K_P} = 2J\sqrt{1-\left(h/J\right)^2}$. Notably, one can arrive at a harmonic oscillator form of $H_{\rm LMG}$ through Holstein--Primakoff transformation as well, for $N \to \infty$ \cite{campbell15shortcut, das2006_infinite}.  

\section{Floquet states}
\label{appFloquet}

\begin{figure}[htbp]
    \centering
    \includegraphics[width=\linewidth]{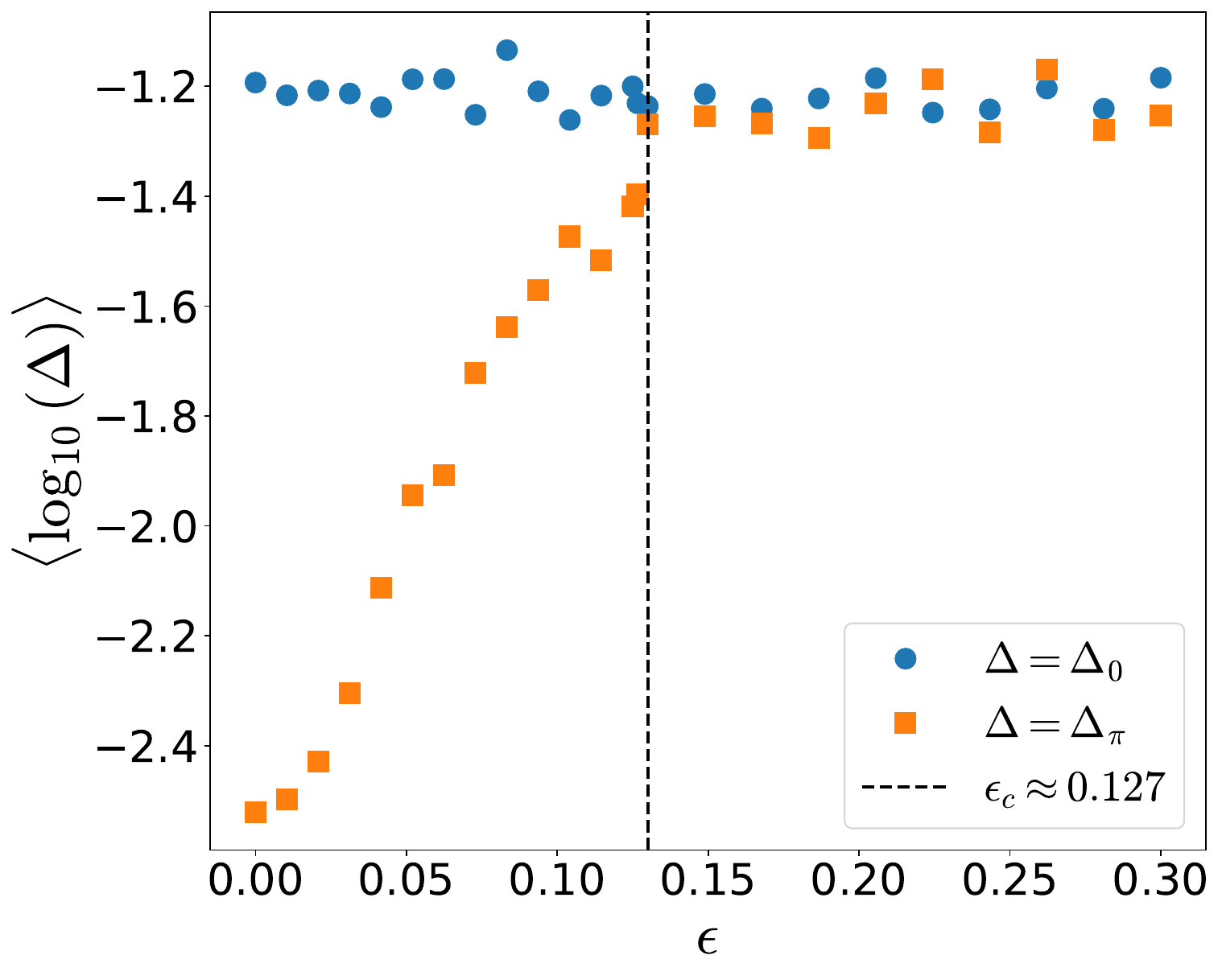}
    \caption{\justifying{Plot of $\langle \log_{10}\Delta\rangle$ with $\epsilon$. Here $N=100,  h=0.3$, and $\tau=0.6$.  }} 
    \label{fig:floquet state}
\end{figure}

As discussed in Refs. \cite{khemani16phase, keyserlingk16phase}, $2$-DTCs are associated with Floquet operators (see Eq. \eqref{eqfloquet}) with pairs of quasienergies $\mu_{\alpha}$ separated by $\pi/\tau$; here $\alpha$ denotes an eigenvalue index, with $\mu_{\alpha^{\prime}} > \mu_{\alpha}$ for $\alpha^{\prime} > \alpha$. Consequently. following \cite{russomanno17floquet}, one can define 
\ba
\Delta_0 &=& \mu_{\alpha + 1} - \mu_{\alpha},\non\\
\Delta_{\pi} &=& \left(\mu_{\alpha + N/2} - \mu_{\alpha}\right) - \pi/\tau.
\ea
The existence of a DTC demands $10^{\langle \log_{10}\Delta_{\pi}\rangle} \to 0$ faster than $10^{\langle \log_{10}\Delta_0 \rangle}$, for $N \to \infty$. As shown in Fig.~\ref{fig:floquet state} , this condition is valid only for $\epsilon \lessapprox 0.127$, thus once more verifying the value of $\epsilon_c$.

\section{Quantum fluctuations}
\label{AppB}

\begin{figure}[htbp]
    \centering
     \includegraphics[width=\linewidth]{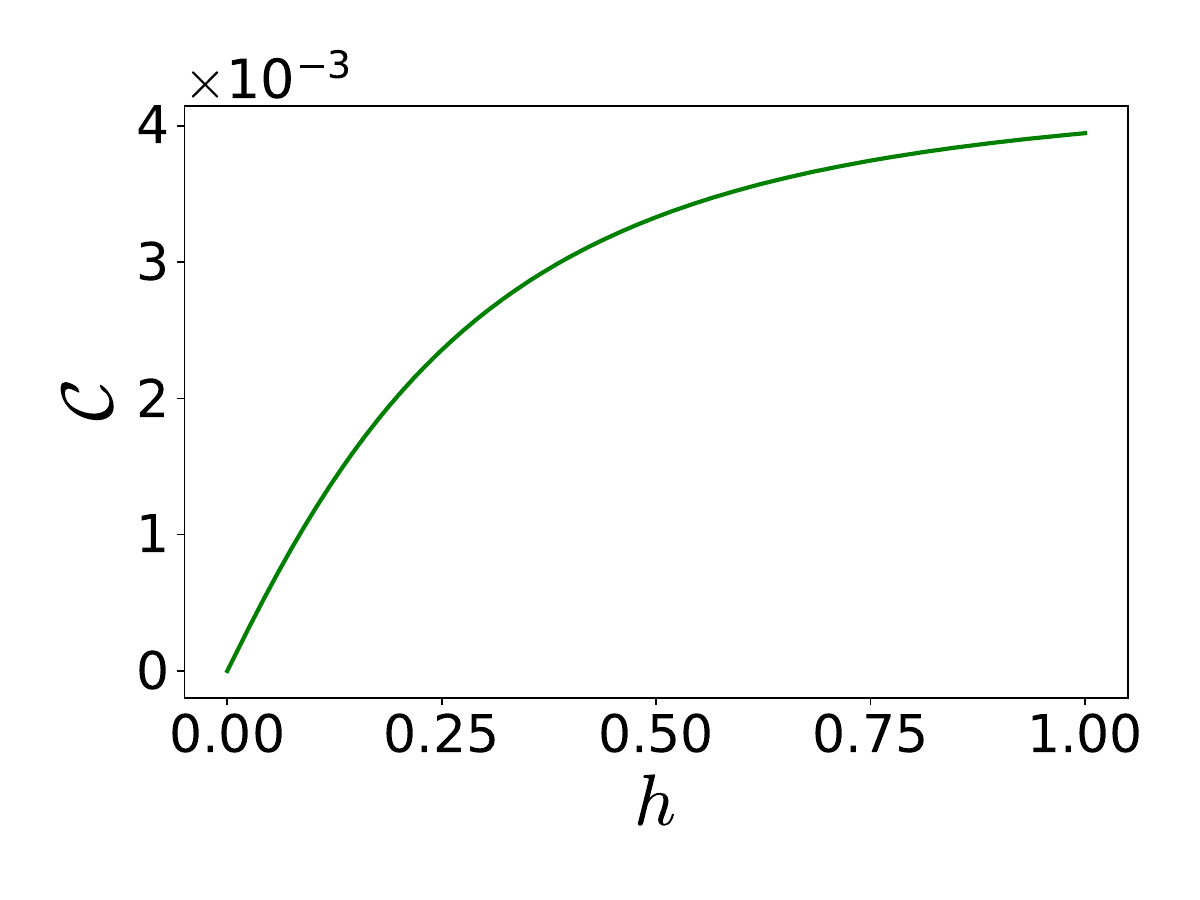}
    \caption{\justifying{Growth of  $\mathcal{C}$ with transverse field strength $h$. Other parameters are $N=100$ and $\tau=0.6$}}
    \label{fig:non-commuting}
\end{figure}
In order to quantify the effect of the transverse magnetic field on the stability of the time-crystalline phase in finite-size systems, we introduce the measure
\[
\mathcal{C}(h) \;=\; \frac{1}{N} \frac{\bigl\| \,[H_1, \, h S_x] \,\bigr\|}{\|H_1\|},
\quad \text{where }
\|A\| \;=\; \operatorname{Tr}\!\left( \sqrt{ A^\dagger A } \right).
\]
Here, the numerator quantifies the quantum fluctuations arising due to non-commutation of $H_1$ and $S_x$, while the denominator provides a normalization with respect to  $H_1$ itself and the system size $N$. As shown in Fig. \ref{fig:non-commuting}, in the regime considered here, $\mathcal{C}(h)$ increases with increasing $h$, which also corresponds to decreasing values of $\epsilon_c$ in Figs. \ref{fig phase diagram}(a) and \ref{fig phase diagram}(b).

\end{document}